\documentclass[review]{elsarticle}

\usepackage{lineno,hyperref}
\modulolinenumbers[5]

\journal{Journal of \LaTeX\ Templates}

\def\Journal#1#2#3#4{{#1} {#2} (#4) #3 }

\def\ARNPS{{\em Ann. Rev. Nucl. Part. Sci.}}

\def\EJP{{\em Eur. J. Phys.}}
\def\EPJA{Eur. Phys. J. A}
\def\IJMPE{International Journal of Modern Physics E}

\def\JCP{Journal of Computational Physics}

\def\JPG{{\em J. Phys. G: Nucl. Part. Phys.}}
\def\NATURE{Nature}

\def\NPA{{\em Nucl. Phys.} A}

\def\PRC{Phys. Rev. C}
\def\PRD{{\em Phys. Rev.} D}

\def\PLB{{\em Phys. Lett.} B}

\def\PRL{\em Phys. Rev. Lett.}

\def\PREV{\em Phys. Rev.}
\def\PREP{ Physics Reports}
\def\PROG{Progress in Particle and Nuclear Physics}

\usepackage{color}
\usepackage{float}

\begin{document}

\begin{frontmatter}

%\title{Elsevier \LaTeX\ template\tnoteref{mytitlenote}}
\title{Neutron stars and supernova explosions in the framework of Landau's theory}


\author[mymainaddress,mysecondaryaddress]{H. Zheng\corref{mycorrespondingauthor}}
\cortext[mycorrespondingauthor]{Corresponding author}
\ead{zheng@lns.infn.it}
\author[mymainaddress,reuaddress]{J. Sahagun}
\author[mymainaddress,mythirdaddress]{A. Bonasera}

\address[mymainaddress]{Cyclotron Institute, Texas A\&M University, College Station, TX 77843, USA}
\address[mysecondaryaddress]{Physics Department, Texas A\&M University, College Station, TX 77843, USA}
\address[reuaddress]{Department of Physics, University of California, Berkeley, CA 94720, USA}
\address[mythirdaddress]{Laboratori Nazionali del Sud, INFN, via Santa Sofia, 62, 95123 Catania, Italy}

\begin{abstract}
A general formula of the symmetry energy for many-body interaction is proposed and the commonly used two-body interaction symmetry energy is recovered. Within Landau's theory (Lt), we generalize two equations of state (EoS) CCS$\delta$3 and CCS$\delta$5 to asymmetric nuclear matter. We assume that the density and density difference between protons and neutrons divided by their sum are order parameters. We use different EoS to study neutron stars by solving the TOV equations. We demonstrate that different EoS give different mass and radius relation for neutron stars even when they have exactly the same ground state (gs) properties ($E/A$, $\rho_0$, $K$, $S$, $L$ and $K_{sym}$). Furthermore, for one EoS we change $K_{sym}$ and fix all the other gs parameters. We find that for some $K_{sym}$ the EoS becomes unstable at high density even for neutron matter. This suggests that a neutron star (NS) can exist below and above the instability region but in different states: a quark gluon plasma (QGP) at high density and baryonic matter at low density. If the star's central density is in the instability region, then we associate these conditions to the occurrence of Supernovae (SN).
\end{abstract}

\begin{keyword}
EoS \sep Neutron star \sep Supernova \sep Landau's theory
%\texttt{elsarticle.cls}\sep \LaTeX\sep Elsevier \sep template
%\MSC[2010] 00-01\sep  99-00
\end{keyword}

\end{frontmatter}

%\linenumbers

\section{Introduction}
Recently, a two-solar-mass (1.97$\pm$0.04 $M_\odot$) neutron star (NS)  PSR J1614-2230 has been observed using Shapiro delay \cite{twomsun, lattimerprl05, lattimer12}. This new discovery provided a limit for heavy neutron stars and constrains the nuclear equation of state (EoS). M. Dutra {\it et al.} have done a systematic study of 240 Skyrme EoS using the mass of this particular NS as one of the criteria to select the EoS. They found that only five out of 240 EoS pass all the criteria \cite{stone2012}. We also studied the correlation of the maximum mass and radius of NS with the incompressibility $K$, the symmetry energy $S$, the slope of the symmetry energy $L$ and its incompressibility $K_{sym}$ by solving the TOV equations for pure neutron matter (PNM) with 159 Skyrme EoS \cite{huappnp}. We found that there might be a correlation between the maximum mass and the radius of the NS with $K_{sym}$. There are many factors that can affect the maximum mass-radius relation of NS, e.g. the proton fraction in the NS \cite{huappnp}, hyperons \cite{agrawal2012, lonardoni2014}, three-neutron interactions \cite{gandolfi2012}, the hadron-quark phase transition (PT) \cite{shao2011, logoteta2013, NSmw2}, the strong magnetic field \cite{dexheimer2012}. Thus, it is difficult to constrain the EoS without taking into account all the observations of the NS. In this paper, we discuss two EoS derived from Landau's theory (Lt): CCS$\delta$3 and CCS$\delta$5 \cite{huavirial} and two simple Skyrme EoS: CK225 and CK225$_1$ \cite{huappnp, huavirial}, which, in principle, do not include any PT at high densities. We first show that even though the different EoS might have the same properties, i.e. same $E/A$, ground state density $\rho_0$, $K$, $S$, $L$, $K_{sym}$, for the ground state (gs) of symmetric nuclear matter, they result in completely different mass-radius relations for NS. Thus fixing the parameters entering the EoS on the gs of symmetric nuclear matter is not sufficient to make predictions for NS. The maximum mass of the NS is the result of the competition between a possible PT and the highest power of density in the EoS \cite{huappnp}. The CCS$\delta$5 EoS contains enough free parameters in such a way that we can fix all of them to currently accepted values apart $K_{sym}$. We show that by changing this last value we find solutions which are unstable at high densities. In Lt this is the result of a first order PT. The resulting scenario is that very massive neutron stars are probably in the quark phase (Lt does not tell us what species are involved, but only that there is a PT) while lighter stars are made of baryons. For intermediate systems, the stars are unstable and we associate these to the occurrence of supernova (SN) explosions.  Other EoS might display instabilities at high densities as well if one opportunely changes the $L$ parameter for a given symmetry energy $S$.

\section{The Nuclear Equation of State}
In Lt, the free energy or EoS at zero temperature can be expanded as a function of one or more order parameters \cite{huangstat}. In ref. \cite{huavirial}, we have proposed the EoS of symmetric nuclear matter assuming $\rho$ is an order parameter:
\begin{equation}
\frac{E}{A} =  \tilde \varepsilon_f \tilde \rho^{2/3} + \sum_{i=1}^n \frac{A_i}{i+1}\tilde \rho^i, \label{eosvirial}
\end{equation}
where $\tilde \varepsilon_f  = 22.5$ MeV is the average Fermi energy and $A_i$ are coefficients. $\tilde \rho = \frac{\rho}{\rho_0}$ and $\rho_0=0.16$ fm$^{-3}$ is the gs density. The first term on the right hand side of Eq. (\ref{eosvirial}) is the (non relativistic) kinetic energy of a free Fermi gas which is unimportant at high densities. The other terms in the sum of Eq. (\ref{eosvirial}) are the contributions from many-body interactions. The different $n$-cutoff in Eq. (\ref{eosvirial}) results in different EoS and different number of parameters to be fixed from known properties.  Notice that Eq. (\ref{eosvirial}) is also referred to a virial expansion of the EoS in terms of density. The difference is that in Lt a PT is assumed and the density is an order parameter. The Lt does not specify if the PT is from nuclear to quark phase or hyperons, but simply that there is a PT. Depending on the values of the parameters, the PT could be first, second order or a simple cross-over. Experiments or theoretical considerations should give constraints to the parameter values. We use the known properties of symmetric nuclear matter at normal density to determine the coefficients in Eq. (\ref{eosvirial}), i.e. $\left.\frac{E}{A} \right\vert_{\rho=\rho_0}= -15$ MeV, $K = 225$ MeV and $\left. P \right\vert_{\rho = \rho_0}=0$. When the number of coefficients is more than 3, we need more conditions to pin down the coefficients. To do that, we assume there is a PT from nuclear matter to quark gluon plasma (QGP), or in general from phase A to phase B, at high density with the conditions $\left. \frac{\partial^k P}{\partial \rho^k} \right\vert_{\rho = \rho_c} = 0$ where $P$ is the pressure, $k=1, 2 \dots$. These conditons imply a first ($k=1$) or a second ($k=1, 2\dots$) order PT for symmetric nuclear matter \cite{huavirial}. In the case of a first order PT we need to know the critical density otherwise the number of conditions will not be enough to fix the parameters entering the EoS. The choice of a second order PT is just to have enough constraints for the parameters entering Eq. (\ref{eosvirial}). If future experimental data or theoretical considerations will point to a different type of PT and critical density, then the parameters might be adjusted to reproduce those results. When we stop the expansion in Eq. (\ref{eosvirial}) at $n=3$ or $n=5$, we obtain $A_n<0$ which is unphysical since the matter would collapse at high densities.  We obtain two EoS dubbed CCS$\delta$3 and CCS$\delta$5 when $n=4$, O($\rho^4$), and $n=6$, O($\rho^6$), respectively \cite{huavirial}. For these values we have a second order PT for symmetric nuclear matter at different values of the critical density, $\rho_c=2.9354\rho_0$ and $5.2795\rho_0$ respectively. Experimentally, the value of the critical exponent $\delta = 4-5$ is found. The CCS$\delta$3 gives $\delta = 3$ which corresponds to the ``mean field''  or ``classical'' value and CCS$\delta$5 gives $\delta = 5$ closer to experiments  \cite{huangstat, landaustat}. We stress that the assumption of a second order PT is for simplicity since it gives enough constraints to determine the values of the parameters entering Eq. (\ref{eosvirial}) and the critical density. A first order PT requires the knowledge at least of the value of the critical density, thus we need experimental constraints or Lattice QCD calculations at high baryonic densities and zero temperature to fix the parameters.

We can generalize Lt for asymmetric nuclear matter. For instance, we can define the total symmetry energy as 
\begin{equation}
E_{s}^{total} = \frac{C_n}{\rho_0^{n}}\int (\rho_n-\rho_p)^{n+1} d^3r, \label{sdef}
\end{equation}
where $n$ is a constant connecting with the many-body interaction and $C_n$ is the corresponding symmetry energy per nucleon. For infinite nuclear matter, i.e. assuming a constant density difference, we have 
\begin{eqnarray}
E_s^{total} &=& \frac{C_n}{\rho_0^{n}} (\rho_n-\rho_p)^{n+1} V\nonumber\\
&=&  \frac{C_{n}}{\rho_0^{n}} (\rho_n-\rho_p)^{n+1} \frac{A}{\rho} \nonumber\\
&=& A C_{n} m_\chi^{n+1} \tilde \rho^{n},
\end{eqnarray}
where $A$ is the number of nucleons, $m_\chi = \frac{\rho_n-\rho_p}{\rho}$ is the asymmetry order parameter \cite{morder1, morder2}. Then the symmetry energy per nucleon is 
\begin{equation}
\frac{E_s^{total} }{A} = C_{n} m_\chi^{n+1} \tilde \rho^{n}. \label{sperAdef}
\end{equation}
When $m_\chi = 1$ and $\rho=\rho_0$, the symmetry energy per nucleon is $C_n$. If only two-body interactions are important, i.e. $n=1$, the symmetry energy per nucleon is
\begin{equation}
\frac{E_s^{total} }{A} = C_1 m_\chi^{2} \tilde \rho. \label{sperA}
\end{equation}
Thus, the usually used form of the symmetry energy is recovered with the definition Eq. (\ref{sdef}) \cite{comd2001, comd2005, comd2009}.  Notice that the asymmetry order parameter enters to a higher power than the density order parameter, we will see some consequences of this later.

Adopting the symmetry energy per nucleon from Eq. (\ref{sperAdef}), we can generalize the EoS in Lt Eq. (\ref{eosvirial}) for asymmetric nuclear matter:
\begin{equation}
\frac{E}{A}(\rho, m_\chi) =  (1+\frac{5}{9}m_\chi^2)\tilde \varepsilon_f \tilde \rho^{2/3} + \sum_{i=1}^n \frac{A_i}{i+1}(1+c_i m_\chi^{i+1})\tilde \rho^i, \label{gccs3}
\end{equation}
where $C_i = \frac{A_i}{i+1}c_i$ and $c_i$ are unknowns. Assuming that the symmetry energy is symmetric under the exchange of protons and neutrons, results in the odd power terms of $m_\chi$ equal to zero. Therefore, the coefficients $c_i=0$ when $i$ is even, i.e. $c_2 = c_4 = 0$ for CCS$\delta$3 and $c_2=c_4=c_6=0$ for CCS$\delta$5. Thus the highest density term which gives repulsion at high densities is independent of the asymmetry correction.

From Eq. (\ref{gccs3}), we can easily obtain the symmetry energy $S$, the slope of the symmetry energy $L$ and the incompressibility of the symmetry energy $K_{sym}$ \cite{huappnp, baoanrep2008} 
\begin{equation}
S = \left. S(\rho) \right\vert_{\rho=\rho_0}= \left. \Big [\frac{\frac{E}{A}(\rho, 1)+\frac{E}{A}(\rho, -1)}{2} - \frac{E}{A}(\rho, 0)\Big ]\right\vert_{\rho=\rho_0}=\frac{5}{9}\tilde\varepsilon_f +\sum_{i=1, i \in odd}^{n} \frac{A_i}{i+1}c_i, \label{ccseq1}
\end{equation}
\begin{equation}
L = \left. 3\rho_0 \frac{\partial S(\rho)}{\partial \rho} \right\vert_{\rho=\rho_0} = 3\times (\frac{10}{27}\tilde\varepsilon_f+\sum_{i=1, i \in odd}^{n} \frac{iA_i}{i+1}c_i), \label{ccseq2}
\end{equation}
\begin{equation}
K_{sym} = \left. 9\rho_0^2 \frac{\partial S(\rho)}{\partial \rho} \right\vert_{\rho = \rho_0} = 9\times [-\frac{10}{81}\tilde \varepsilon_f+\sum_{i=1, i \in odd}^{n} \frac{i(i-1)A_i}{i+1}c_i]. \label{ccseq3}
\end{equation}
In the CCS$\delta$3 case, we have three quantities $S$, $L$ and $K_{sym}$ and two unknowns $c_1$ and $c_3$. This means that the $S$, $L$ and $K_{sym}$ are correlated  and it is sufficient to fix two of them. In the CCS$\delta$5 case, we have three unknowns $c_1$, $c_3$ and $c_5$, thus we can change $S$, $L$ and $K_{sym}$ independently to study the EoS, i.e. we need more conditions to fix the EoS.  We stress that this simple result has been obtained by invoking invariance of the nuclear force in isospin space. If this invariance is violated, and it might be violated, then we need to add the even terms in the symmetry energy, Eq. (\ref{gccs3}). 

Skyrme interactions are widely used in the literature \cite{baoanrep2008, skyrmebrown,  baranrep2005}. For instance a simple EoS is CK225 defined in \cite{huappnp}
\begin{equation}
\frac{E}{A}(\rho, m_{\chi}) = (1+\frac{5}{9}m_{\chi}^2) \tilde \varepsilon_f \tilde \rho^{2/3} + (1+c_1m_{\chi}^2)\frac{A_1}{2}\tilde \rho + (1+c_2 m_{\chi}^2) \frac{A_2}{1+\sigma} \tilde \rho^\sigma, \label{ck225}
\end{equation}
where $A_1=-210.0$ MeV, $A_2= 157.5$ MeV and $\sigma = \frac{4}{3}$. In this EoS, all the many-body (more than two-body) interactions have been absorbed into the term $\tilde \rho^\sigma$. The symmetry energy is approximated to the first even term. When $c_2=0$, this EoS becomes the one used in the constrained molecular dynamics model (CoMD) \cite{comd2001, comd2005, comd2009} and we dub it as CK225$_1$. The corresponding physical quantities $S$, $L$ and $K_{sym}$ are
 \begin{equation}
 S=\frac{5}{9} \tilde\varepsilon_f + \frac{A_1}{2}c_1 +  \frac{A_2}{1+\sigma}c_2, \label{ck225eq1}
 \end{equation}
 \begin{equation}
  L = 3\times( \frac{10}{27}\tilde\varepsilon_f + \frac{A_1}{2}c_1 + \frac{A_2\sigma}{1+\sigma} c_2 ), \label{ck225eq2}
 \end{equation}
 \begin{equation}
 K_{sym} = 9\times [-\frac{10}{81}\tilde\varepsilon_f + c_2 \frac{A_2 \sigma(\sigma-1)}{1+\sigma} ]. \label{ck225eq3}
 \end{equation}
Similar to CCS$\delta$3, only two of the $S, L$ and $K_{sym}$ are independent. In the CK225$_1$ case, we can change $S$ or $L$ since we have one unknown $c_1$ and the resulting $K_{sym} = -25$ MeV is a constant. Incidentally this is the value of the symmetry incompressibility of a free Fermi gas \cite{huappnp}. Notice that if the interaction is attractive in the asymmetry part, then $K_{sym}$ will be lower than the free Fermi gas value, it will be higher if the interaction is repulsive. In order to get a repulsive term at high densities, the last term in Eq. (\ref{ck225}) must be positive, i.e. $1+c_2\ge 0$ for PNM. Solving Eqs. (\ref{ck225eq1}, \ref{ck225eq2}, \ref{ck225eq3}) for $c_2=-1$ and fixing $S=28.5$ MeV for instance, gives $L=5.5$ MeV and $K_{sym}=-295$ MeV which must be considered as the lowest limits for those quantities and for this EoS.

  \begin{figure} [H]  
        \centering
   %     \begin{tabular}{c}
        \includegraphics[scale=0.5]{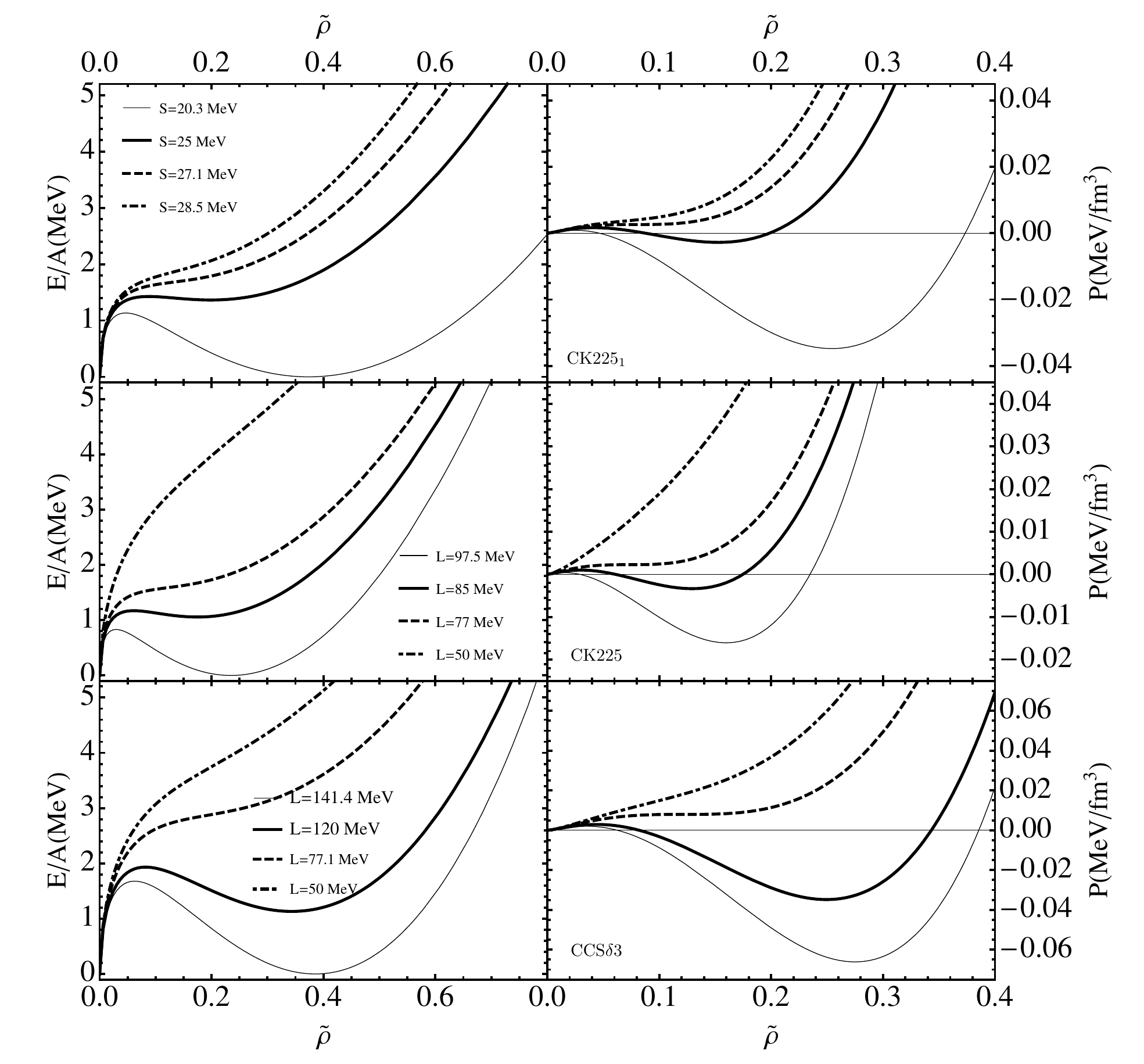}  
     %   \end{tabular}
\caption{The energy per nucleon and pressure versus density for CK225$_1$, CK225 ($S=28.5$ MeV) and CCS$\delta$3 ($S=28.5$ MeV) for PNM. Densities below $\rho_0$ only are plotted. The thin black line is obtained imposing $E/A=0$ MeV at $\rho_x$, and $P|_{\rho=\rho_x}=0$, i.e. a stable configuration; the thick black line refers to a first order PT; the dashed line refers to a second order PT and the dot-dashed line refers to currently accepted values for $S$ and $L$.} \label{nfig1}
    \end{figure}

\section{Properties of the EoS for Pure Neutron Matter}

There is a consensus of the incompressibility for symmetric nuclear matter $K = 250 \pm 25$ MeV from isoscalar giant monopole resonance (ISGMR) \cite{huappnp, anders2013, youngblood1, youngblood2, piekarewicz1, chen1, cao1} and we use $K=225$ MeV for the remainder of this paper. The symmetry energy $S = 30\pm 5$ MeV and $L=50\pm 40$ MeV. There is no constraint in $K_{sym}$ \cite{huappnp, Sbetty2014}.  

First let us study some properties for the EoS for PNM starting from the simplest one, CK225$_1$ which contains one free parameter only and we choose it to be the symmetry energy $S$. This EoS displays a liquid-gas PT at low density and for symmetric matter. The fate of the PT and/or of the gs of the system depends on the value of $S$. We can impose that the energy per nucleon has a minimum at a density $\rho_x$. Since this introduces a new quantity, i.e. the density of the minimum, we further impose for illustration that $E/A|_{\rho=\rho_x}=0$ MeV and $P|_{\rho=\rho_x}=0$ for PNM. Using these conditions, we get $S=20.3$ MeV and $\rho_x=0.37 \rho_0$. Such a value is outside the range of currently accepted $S$. If we instead assume the occurrence of a PT, the first derivative of the pressure must be zero. This will occur at a critical density which is unknown. For a second order PT, the second derivative of the pressure is also zero and using these two conditions we can fix both the value of $S$ and the critical density for PNM. In Fig. \ref{nfig1}, top, we plot the $E/A$ (left panel) and the pressure (right panel) vs density for different values of $S$. The curve $S=20.3$ MeV and $L$=48.4 MeV gives a minimum of the $E/A$ for PNM, i.e. for such $S$ a NS would be self-bound, there is no need for the gravitational force to bound the star. There is no experimental evidence  that two or more neutrons are bound. Of course if such a system would exist with a minimum at $E/A$=0 MeV, it would be unstable and quickly some neutrons will decay into protons in free space. Furthermore the derived value for $S$ is outside the accepted values from mass-formula and other studies \cite{huappnp}. A first order PT occurs for $S=25$ MeV and $L$=62.5 MeV, currently accepted values. Increasing the values of $S$ to 27.1 MeV and $L$=68.9 MeV results in a second order PT. These cases are the remnant of the liquid-gas PT for PNM.  Higher values of $S$ give a monotonic increase of pressure vs density.

  \begin{figure} [H]  
        \centering
   %     \begin{tabular}{c}
        \includegraphics[scale=0.5]{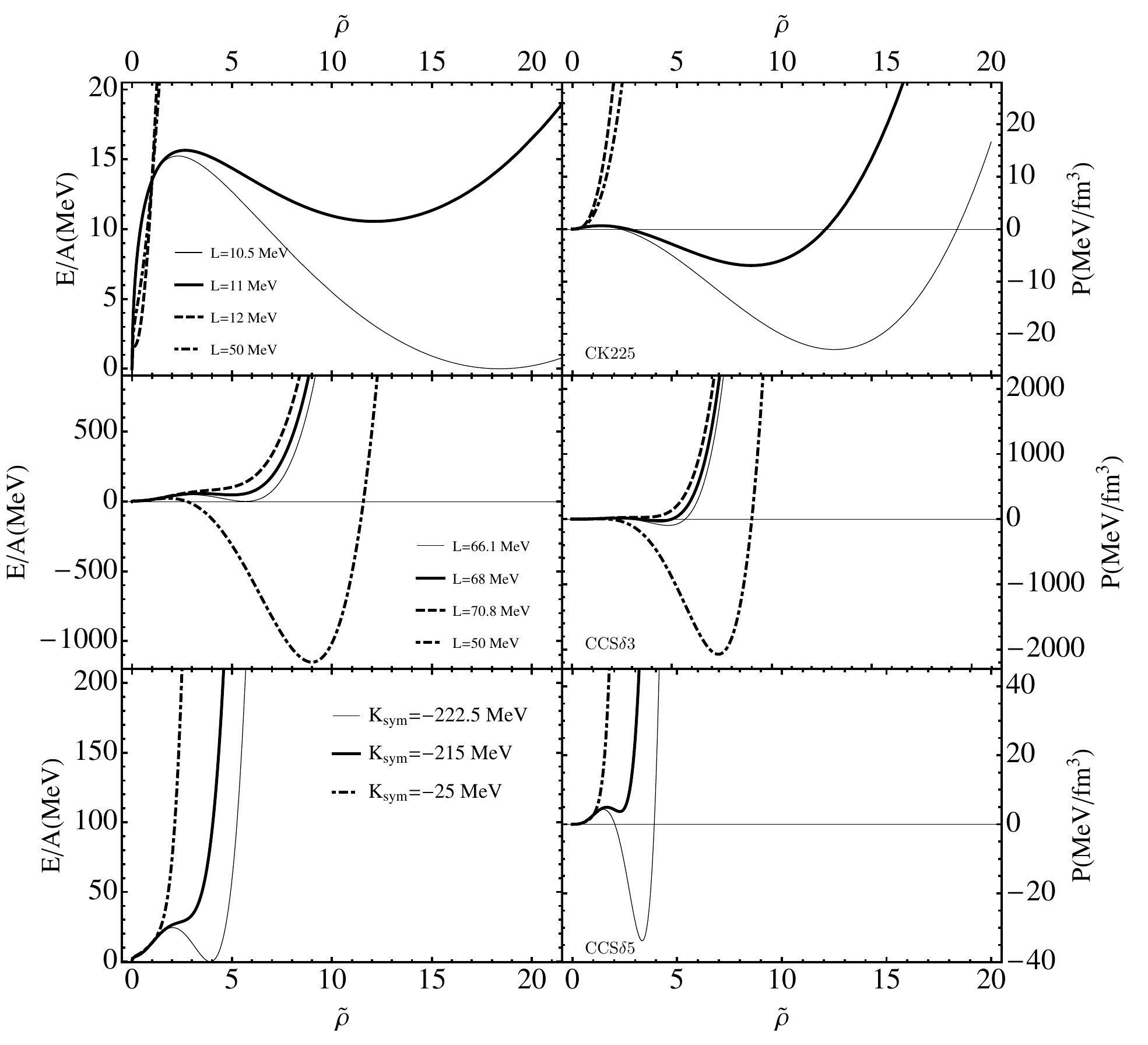}  
     %   \end{tabular}
\caption{The energy per nucleon and pressure versus density for CK225, CCS$\delta$3 and CCS$\delta$5. Solutions for densities above the gs one are plotted and $S=28.5$ MeV. The thin black line refers to $E/A=0$ MeV at $\rho_x$, the thick black line refers to a first order PT, the dashed line refers to second order PT and the dot-dashed line refers to some typical values. The $L$=50 MeV for CCS$\delta$3 is unphysical. There is no solution for second order PT for CCS$\delta$5 since we require $E/A>$-16 MeV in all the density region. } \label{nfig2}
    \end{figure}
    
The CK225 and CCS$\delta$3 EoS depend on two parameters which we assume to be $S$ and $L$. The behavior of these EoS is similar to what we have discussed before for each value of $S$ and changing $L$. Let us assume $S=28.5$ MeV for these EoS. In Fig. \ref{nfig1} (middle panel) we plot $E/A$ (left panel) and $P$ (right panel) vs $\rho$ for different values of $L$ corresponding as before to a bound state of PNM, a first or a second order PT. In Fig. \ref{nfig2}, similar results are reported but for densities higher than the gs density of symmetric nuclear matter. All the EoS, apart CK225$_1$, admit solutions both at low and high densities. The CCS$\delta$5 EoS, assuming $S=28.5$ MeV as well,  has a solution for low densities, but with a very pronounced minimum in the binding energy, thus it is unphysical and not reported in Fig. \ref{nfig1}. The result for CCS$\delta$3 $L=50$ MeV, which is within currently accepted values \cite{huappnp}, is also given for reference. It gives a minimum in the energy deeper than the gs of symmetric nuclear matter, thus it is unphysical, see Fig. \ref{nfig2}, middle panel. For the same EoS and $L>66$ MeV we get respectively a bound state ($E/A=0$ MeV at the minimum), a first order or a second order PT at high densities, Fig. \ref{nfig2}. These are perfectly acceptable values of $L$ and in some cases one can get a minimum or a PT even at low densities, see Fig. \ref{nfig1}. Thus we have a very large variety of situations, with the possibility of a PT at low (liquid-gas) or high (QGP) densities. We stress that those situations never occur for exactly the same value of $L$, thus a good precision measurement of this parameter is needed in order to restrict the large variety of physical situations. A similar variety of situations is possible also for the `simple' CK225 EoS, just by changing the value of $L$ and fix $S=28.5$ MeV, see Figs. \ref{nfig1} and \ref{nfig2}. However, the values of $L$ of interest for the different EoS are not the same. 

The $K_{sym}$ is not independent of $S$ and $L$ for the previous EoS but it is for CCS$\delta$5. For this EoS we fix $S=28.5$ MeV and $L=50$ MeV and vary $K_{sym}$. Of course the situation is quite rich in this case and one can vary $L$ or $S$ as well within accepted values and in so doing all the situations discussed previously might be recovered for this particular EoS. As we see in Fig. \ref{nfig2} (bottom panel) we can get a bound state for $K_{sym}=-222.5$ MeV all the way to completely repulsive pressure for $K_{sym}=-25$ MeV. For intermediate values of $K_{sym}$ we can get a first order PT but not a second order PT. Since we have no idea on what $K_{sym}$ should be, we will put some constraints in the following from NS properties. 

\section{Properties of Neutron Stars}
Let us  fix $K=225$ MeV, $S=28.5$ MeV, and  $L=73$ MeV from CK225$_1$ for illustration. Of course CK225 and CK225$_1$ are the same with $c_1=-0.152381, c_2=0$. Similarly, for CCS$\delta$3, we obtain $c_1=-0.233764, c_3=0$ and $K_{sym}=-25$ MeV as well. Thus CK225$_1$, CK225 and CCS$\delta$3 have the same values for $K$, $S$, $L$ and $K_{sym}$. This motivates us to fix the same values of $S$, $L$ and $K_{sym}$ for CCS$\delta$5 which gives $c_1=-0.231941, c_3=c_5=0$. Notice that only the $m_\chi^2$ term survives in all the EoS, however the density dependence or the EoS is different and as we will show this will produce different results. We would like to stress that these EoS do not show any PT at high densities for neutron matter, see also Figs. \ref{nfig1} and \ref{nfig2} for reference. 

  \begin{figure} [H]  
        \centering
   %     \begin{tabular}{c}
        \includegraphics[scale=0.5]{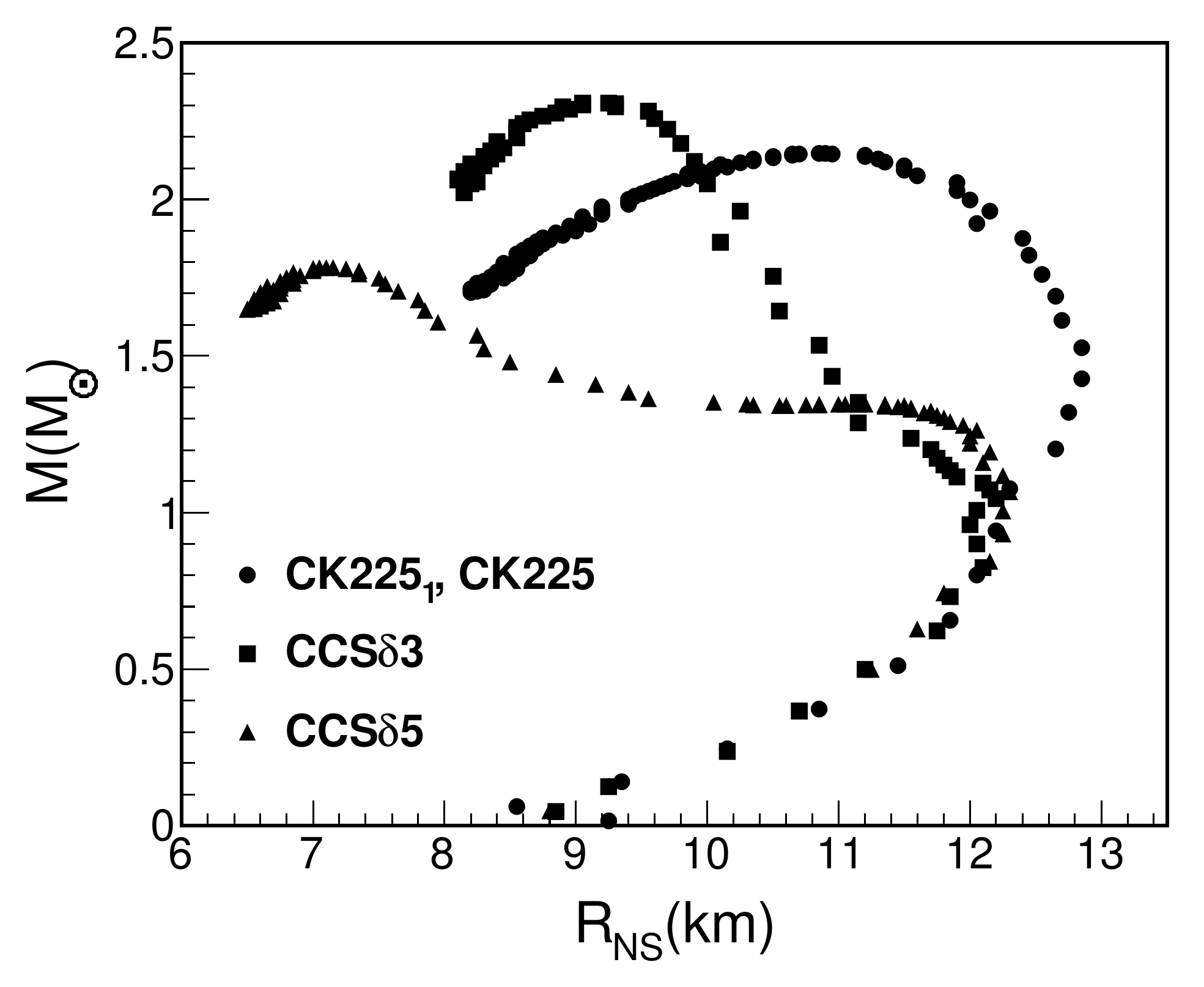}  
     %   \end{tabular}

\caption{The NS mass-radius relation for CK225$_1$, CK225, CCS$\delta$3 and CCS$\delta$5 with same values of $K=225$ MeV, $S=28.5$ MeV, $L=73$ MeV and $K_{sym} = -25$ MeV. For these parameters choice, CCS$\delta$5 is excluded by the observations \cite{twomsun}.}  \label{nfig3}
    \end{figure}

For the four EoS, CK225$_1$, CK225, CCS$\delta$3 and CCS$\delta$5, we solve the TOV equations for the PNM to obtain the mass-radius relation of neutron stars \cite{tov1, tov2, svtov1, svtov2, svtov3, svtov4}. The results are shown in Fig. \ref{nfig3}. We can see that the mass-radius relations for the neutron stars are different for the four EoS (CK225$_1$ and CK225 are the same in this case) even though they have the same values of $K, S, L$ and $K_{sym}$. This indicates that the mass-radius relation of the NS is not only determined by $K, S, L$ and $K_{sym}$, but the high density dependence of the EoS is crucial. We also notice that there are wiggles in the mass-radius relation of PNM NS for CCS$\delta$3 and CCS$\delta$5 rather than CK225$_1$ and CK225 because of a cross-over at high density for the two EoS. Thus, the second order PT assumed in deriving CCS$\delta$3 and CCS$\delta$5 becomes cross-over for these parameters choice. The maximum mass of the NS for CCS$\delta$3 is larger than the one for CK225$_1$ or CK225. The reason for this is because of the higher power of $\tilde\rho$ in CCS$\delta$3 compared to CK225. But the maximum mass of the NS for CCS$\delta$5 (which contains even higher power density values) is lower than the one for CCS$\delta$3. We recall that the critical density for symmetric nuclear matter of the PT for CCS$\delta$5 is higher than the one for CCS$\delta$3 \cite{huavirial}. The `missed phase transition' or cross-over softens the EoS and causes the pressure for CCS$\delta$5 to decrease compared to the one for CCS$\delta$3 at the same density. Therefore, the EoS CCS$\delta$5 can support less massive NS than CCS$\delta$3. However, fixing gs parameters from experimental data might not be sufficient to reproduce neutron matter properties: we need constraints at high density.

Second, we study how the $K_{sym}$ will affect the neutron stars to see if there is a correlation of the maximum mass and radius of the neutron stars with $K_{sym}$ \cite{huappnp}. We fix $S = 28.5$ MeV using the same value as before and $L = 50$ MeV which is an accepted value of $L$ \cite{huappnp, Sbetty2014} and change $K_{sym}$ for the CCS$\delta$5 EoS. We solve Eqs. (\ref{ccseq1}, \ref{ccseq2}, \ref{ccseq3}) for each $K_{sym}$ to obtain the coefficients $c_1, c_3$ and $c_5$. We notice that CCS$\delta$3 gives bound neutron matter at high density for such values of $S$ and $L$, resulting in $K_{sym}=-232$ MeV, see Fig. \ref{nfig2}. There is no indication that neutrons can give bound matter, thus this EoS with these parameters is unphysical, i.e. $L>50$ MeV. It is a general property of all the EoS, if we decrease $K_{sym}$, we make the symmetry part more and more attractive until it gives a bound state at some density.  For the particular choice of $S=28.5$ MeV and $L=50$ MeV as in Figs. \ref{nfig1} and \ref{nfig2}, CK225 does not display a minimum in the $E/A$ plot, while CCS$\delta$5 displays a variety of situations. In fact by decreasing $K_{sym}$ we have a completely repulsive EoS, a first order PT and finally a bound state when $K_{sym}=-222.5$ MeV, see Fig. \ref{nfig2}.

%\begin{figure} [H]  
%\centering
%\begin{tabular}{c}
%%                   \includegraphics[scale=0.5]{figure2_a.eps} \\
% %                    \includegraphics[scale=0.5]{figure2_b.eps} \\
% %                    \includegraphics[scale=0.5]{figure2_c.eps}
%   \includegraphics[scale=0.4]{figure2.pdf}
%\end{tabular}
%\caption{(Color online) The top panel is the energy per nucleon, the middle panel is the pressure and the bottom panel is the incompressiblity versus reduced density respectively. The thick dot-dashed line, thin black line, black line and dashed line are for CCS$\delta$5 with $K_{sym}=-222.5$ MeV, $K_{sym} = -215$ MeV, $K_{sym} = -117$ MeV and $K_{sym} = -50$ MeV respectively. The blue long dashed line is for CK225 with $K_{sym} = -117$ MeV and the red dotted line is for CCS$\delta$3 with $K_{sym} = -232$ MeV. [{\color{red} Your comment:  It is hard to see the instability in this plot. Do you have any idea to make this better? The three figures have the same x axis.}] } \label{fig2}
%\end{figure}

\begin{figure} [H]  
\centering
\begin{tabular}{c}
                   \includegraphics[scale=0.5]{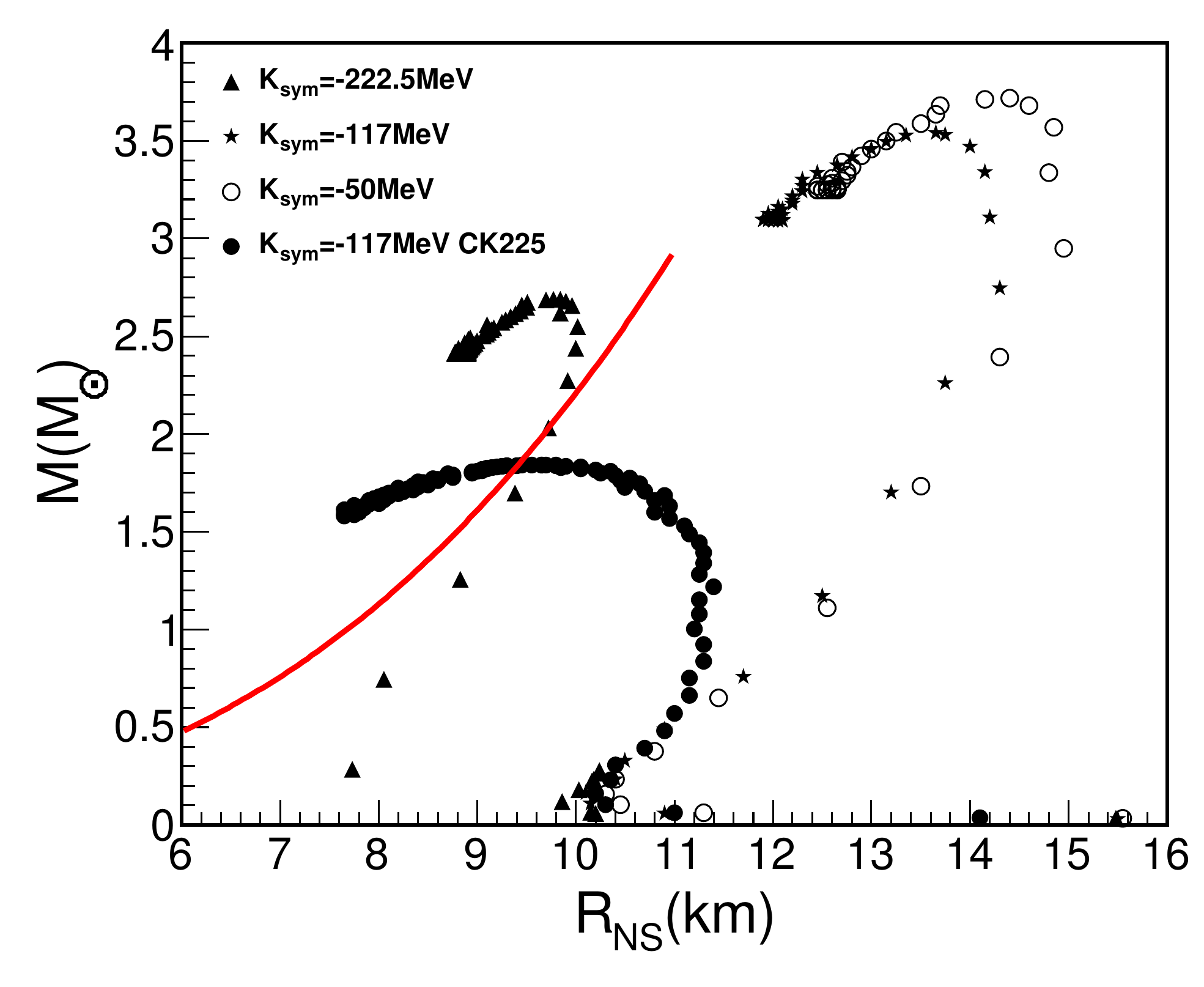}  
\end{tabular}
\caption{(Color online) The NS mass-radius relation for CCS$\delta$5 when we fix $S=28.5$ MeV, $L=50$ MeV and change $K_{sym}$. The results of CK225 are also shown ($K_{sym} = -117$ MeV), which do not reproduce the observations, thus $L>50$ MeV for this EoS. The (red) full line are the estimated results when $K_{sym}=-222.5$ MeV from $M(R_{NS})=\rho_n V m_n$ and they are close to the full TOV results. $\rho_n=3.91\rho_0$ is the PNM `ground state' density for this EoS.} \label{nfig4}
\end{figure}

We solve the TOV equations for each $K_{sym}$ and the results are shown in Fig. \ref{nfig4}. We also show the results of CCS$\delta$5 with the same values of $S$ and $L$ as CK225 for reference in Fig. \ref{nfig4}. We confirm that the mass-radius relations for neutron stars are different for CK225 and CCS$\delta$5 with the same $K$, $S$, $L$ and $K_{sym}$. 

For CCS$\delta$5, it is evident that the maximum mass of the NS increases with $K_{sym}$ as found in ref. \cite{huappnp} using Skyrme type EoS with different values $K$, $S$, $L$ and $K_{sym}$. This is due to the fact that increasing the value of $K_{sym}$ decreases the `softening' of the EoS due to a `missed phase transition' or a cross over. There is an instability region, a first order PT,  for CCS$\delta$5 when $K_{sym}= -215$ MeV. %{\color{red} [$K_{sym}=-215$ MeV??? Since we introduce the maxwell construction. The statement in this paragraph is not really true. You can look at Figs. 4 and 5] [Your comment: = you mean we need the equal sign? Yes. I mean we need to take $K_{sym}=-215$ MeV. Because when $K_{sym}$ is large ennough, we don't have a first order phase transtion any more. Look at the case $K_{sym}=-25$ MeV in Fig. 2.]}

%\begin{figure} [H]  
%\centering
%\begin{tabular}{c}
%                   \includegraphics[scale=0.5]{ksymN228_comp.eps}  
%\end{tabular}
%\caption{The comparison of the results with $K<0$ and $P<0$ as condition to stop the code. } \label{fig2}
%\end{figure}

\begin{figure} [H]  
\centering
\begin{tabular}{c}
                   \includegraphics[scale=0.5]{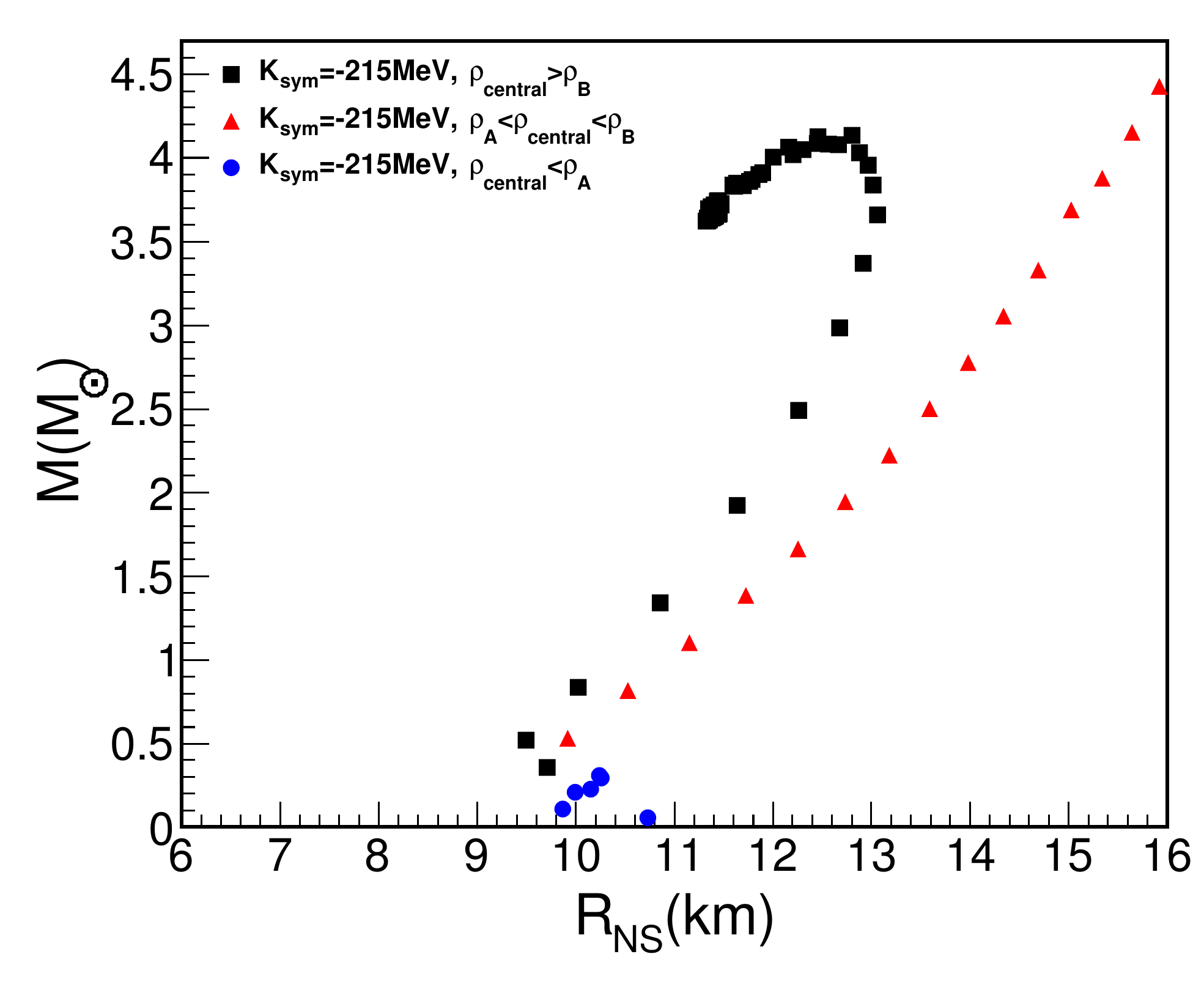}  
\end{tabular}
\caption{(Color online) The NS mass-radius relation for CCS$\delta$5 when we fix $S=28.5$ MeV, $L=50$ MeV and $K_{sym}=-215$ MeV. We performed the Maxwell construction for this case.} \label{nfig5}
\end{figure}

A particular case that we discussed in Fig. \ref{nfig4} is when the EoS has a minimum in the energy per nucleon as function of density for PNM, see also Fig. \ref{nfig2}. It is instructive to see what happens in this `peculiar' case. The NS would be bound without any need of gravitational forces. Knowing the gs density we can calculate the mass simply as: $M(R_{NS})=\rho_n V m_n=\frac{4\pi}{3}R_{NS}^3\rho_n m_n$; where $m_n$ is the neutron mass and $\rho_n=3.91\rho_0$ is the `ground state' density for PNM when $K_{sym}=-222.5$ MeV. Notice that the $E/A$ does not give any contribution to the NS mass since we have chosen the parameters such that it is zero at the minimum, see Fig. \ref{nfig2}.  Such a relation shows a monotonic increase of the NS mass with increasing radius. This increase is not seen in Figs. \ref{nfig3} and \ref{nfig4} apart for small radii. However, the experimental data does not give yet any particular dependence of the mass as function of radius. We can solve the TOV equations for this particular case and the results are reported in Fig. \ref{nfig4}. %{\color{red}IS THIS POSSIBLE? When we solve the TOV equations, the code stops at $\tilde \rho = 3.91221$ where the pressure becomes negative.} 
The result is very close to our estimate and it shows the effect of the gravitational force and relativistic effects.

CCS$\delta$5 displays an instability when $K_{sym}=-215$ MeV corresponding to a first order PT. In this case the TOV equations cannot be solved in the instability region. We have used the Maxwell construction to determine the densities where the two phases (which we call phase A and B) are separated \cite{logoteta2013, NSmw2}. Thus we start the calculations from a high central density $\rho_{central}$ until we reach the end of phase B at density $\rho_B$ and total mass $M_B$. Now we assume that the pressure is approximately constant until we reach the stable new phase A at density $\rho_A$. The TOV equations cannot be solved in the mixed region, thus, in the spirit of the Maxwell construction we assume that the total mass at $\rho_A$ is given by:
\begin{equation}
M_A\approx(2-\rho_A/\rho_B)M_B, \label{maxwm}
\end{equation}
and the radius:
\begin{equation}
R_A\approx(\rho_B/\rho_A)^{1/3} R_B. \label{maxwr}
\end{equation}
Using these as initial conditions for the TOV equations, we can calculate the total mass of the NS. Naturally, when the central density is smaller than $\rho_A$, we have no instabilities. It is interesting to study the cases where the initial density for the TOV equations is very close or inside the unstable region. In the latter case, if the star central density is inside the instability region, phase separation will occur and matter divides into phases A and B at their respective densities. The value of the total mass in the central part of the star might be any, thus we can variate the mass contained in the central region from very high values (say $0.1M_\odot \le M_B \le 4M_\odot$). Knowing the density at point B, for each mass value $M_B$ we can obtain $R_B$ and consequently the values of $M_A$ and $R_A$ using Eqs. (\ref{maxwm}, \ref{maxwr}). Now we can solve the TOV equations starting from the initial conditions of phase A. The resulting mass-radius relation are plotted in Fig. \ref{nfig5}. Starting from a central density of the star above the density of the unstable phase, we can solve the TOV equations and obtain the star density for each value of $R$ until we reach the critical point B. At this point, the TOV solutions become unstable and we use the Maxwell construction through Eqs. (\ref{maxwm}) and (\ref{maxwr}). In this way we determine the stable point A which becomes the new initial conditions for the TOV equations. In this way we are able to obtain the mass-radius relation for the NS represented by the full squares in Fig. \ref{nfig5}. If the central density of NS is below the critical point A, then we can easily solve the TOV equations and the corresponding results are given by the full circles. This solution corresponds to a completely stable NS with one phase only, say baryon matter, with very small mass values. These stars have not been experimentally observed so far probably because their gravitational effects are too small or because they evolve into something else. Of course if they exist in large numbers then they could give a contribution to solve the dark matter `puzzle' \cite{lattimerprl05, mpbook}. An interesting case is when the central density is exactly inside the instability region. For these cases the TOV equations cannot be solved, thus we have to use Eqs. (\ref{maxwm}) and (\ref{maxwr}) as an approximation. Thus we choose different values for $M_B$ within the limits discussed above and determine the conditions for phase A. Starting from these conditions, phase A, we can solve the TOV equations. The results are given in Fig. \ref{nfig5} by the solid triangles and they have been obtained by using the relation M=$\rho_BVm_n$ as we did before for the stable solutions. Rather large values of the mass are obtained and these values have not been confirmed experimentally so far. Naturally, we expect that since these NS have a central density in the unstable region then they might explode, i.e. become SN \cite{NSSN2014}. This will of course be facilitated if there is a second massive star nearby which will break our assumption of the spherical symmetry.

\begin{figure} [H]  
\centering
 \includegraphics[scale=0.5]{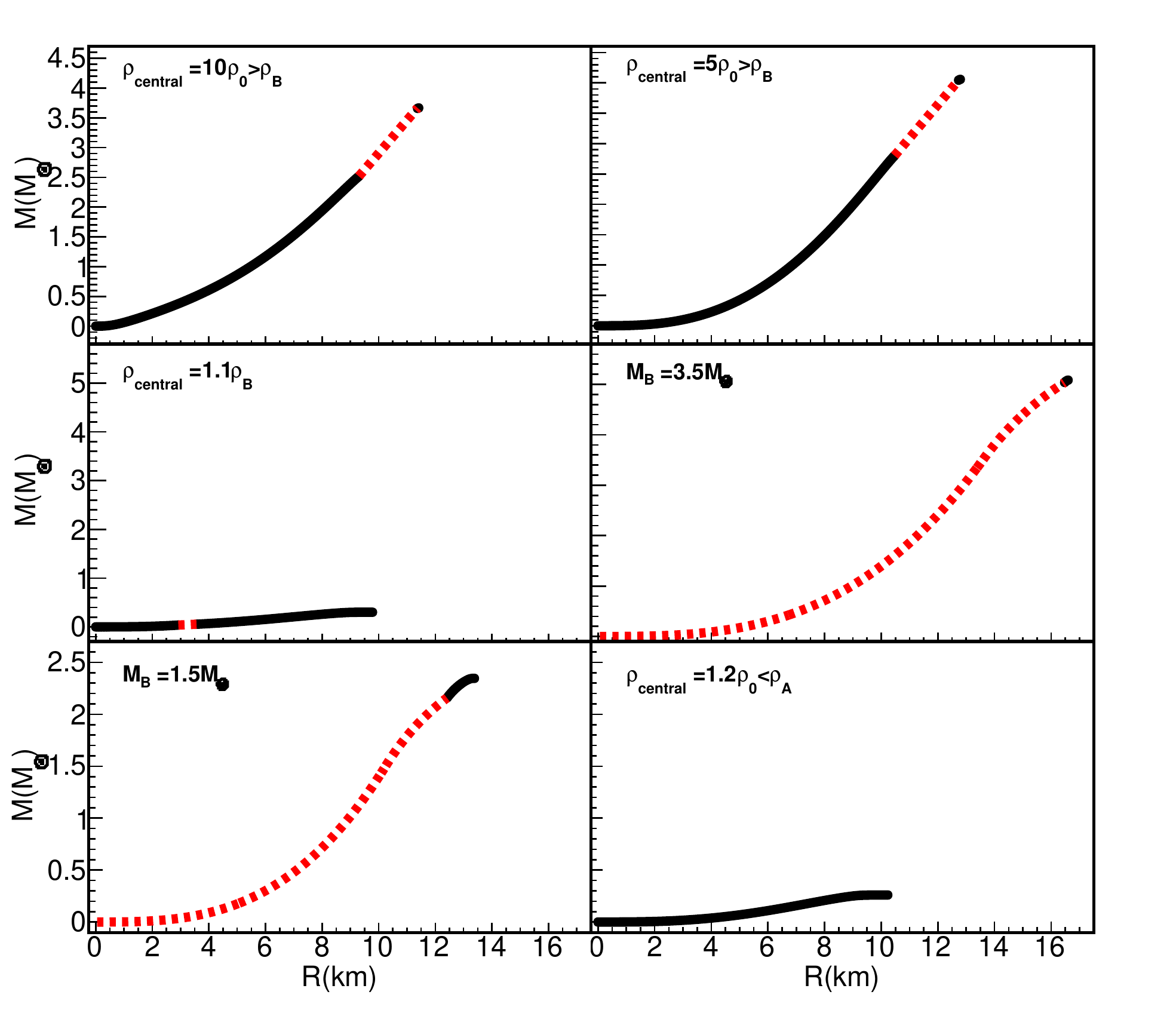}  
\caption{(Color online) The mass versus R for CCS$\delta$5 when $K_{sym}=-215$ MeV for different central density values indicated in the figure. When $\rho_{central}$ is in the instability region (the fourth and fifth panels) we choose $M_B=3.5 M_\odot, 1.5 M_\odot$ respectively. The last panel is obtained when the central density is outside the instability region, thus the baryonic phase only.} \label{nfig6}
\end{figure}

It is instructive to discuss the solutions of the TOV equations separately for each case. In Fig. \ref{nfig6}, we plot the values obtained solving the TOV equations for different initial conditions. The TOV solutions are given by the full lines while the results in the unstable region given by Eqs. (\ref{maxwm}) and (\ref{maxwr}) are given by the dashed lines. The first three panels correspond to central densities above point B. As we see, there is a large part of the star in the unstable region. The stable surface region becomes bigger and bigger, the closer the central density gets to the critical point B. We can argue that for the latter condition the NS is stable since the gravitational force might be able to constrain the unstable region. However, if the stable surface region is too small, or alternatively the mass contained in the unstable region is too large compared to the total NS mass, then part of the star mass might be `evaporated' in order to reduce the  instability region. Of course, from these estimates we cannot say how much mass will be `evaporated' and detailed hydrodynamical calculations are needed using these EoS. Since the observations (so far) give a maximum NS of about $2.5\pm0.5 M_\odot$ \cite{lattimerprl05, lattimer12}, we can argue that for the two top panels cases in Fig. \ref{nfig6}, the unstable region will be `evaporated' or `blown away' which will bring to a rearrangment of the NS with new values of its mass. Of course, these probably are more the conditions for SN explosions.

\begin{figure} [H]  
\centering
\begin{tabular}{c}
                   \includegraphics[scale=0.5]{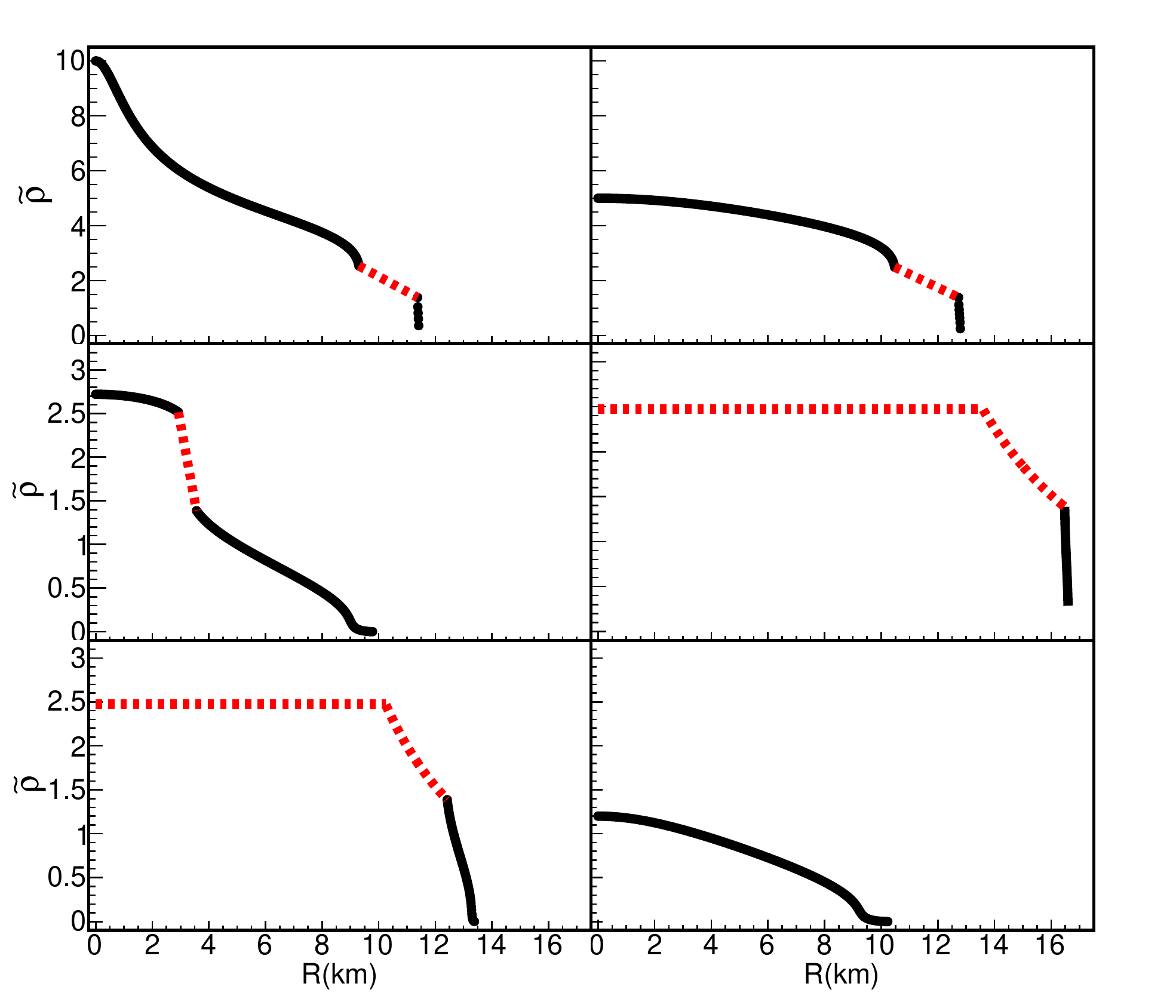}  
\end{tabular}
\caption{(Color online) The density versus R for CCS$\delta$5 when $K_{sym}=-215$ MeV when $\rho_{central}= 10\rho_0, 5\rho_0,1.1 \rho_B$ (first three panels) %. when $\rho_{central}>\rho_B=2.474\rho_0$ 
and $M_B=3.5 M_\odot, 1.5 M_\odot$ when $\rho_{central}$ is in the instability region (the fourth and fifth panels). $\rho_{central}<\rho_A=1.387\rho_0$, i.e. baryonic matter, is plotted in the sixth panel.} \label{nfig7}
\end{figure}

The two following panels in Fig. \ref{nfig6} correspond to a central density of the NS inside the instability region. The corresponding mass-radius relation is plotted in the figure. The solution in the unstable region from Eqs. (\ref{maxwm}) and (\ref{maxwr}) is given by the dashed line while the full line is obtained from the TOV equations starting from phase A. We expect these cases to correspond to `explosive' events and the remanent part of the star will be very small.  Finally, the stable solution of the TOV equations are found when the central density is below the critical point A. For these cases we have only one phase. We stress that the particular values discussed above are obtained for the particular choice of $S$ and $L.$ Other choices will result in different critical densities for phases A and B. It is very important to pin down the values of these two parameters to a very high precision. Experiments in heavy ion collisions also using radioactive nuclei might be the only tools we have in terrestrial laboratories to constrain the parameters entering the EoS. 

\begin{figure} [H]  
\centering
\begin{tabular}{c}
                   \includegraphics[scale=0.5]{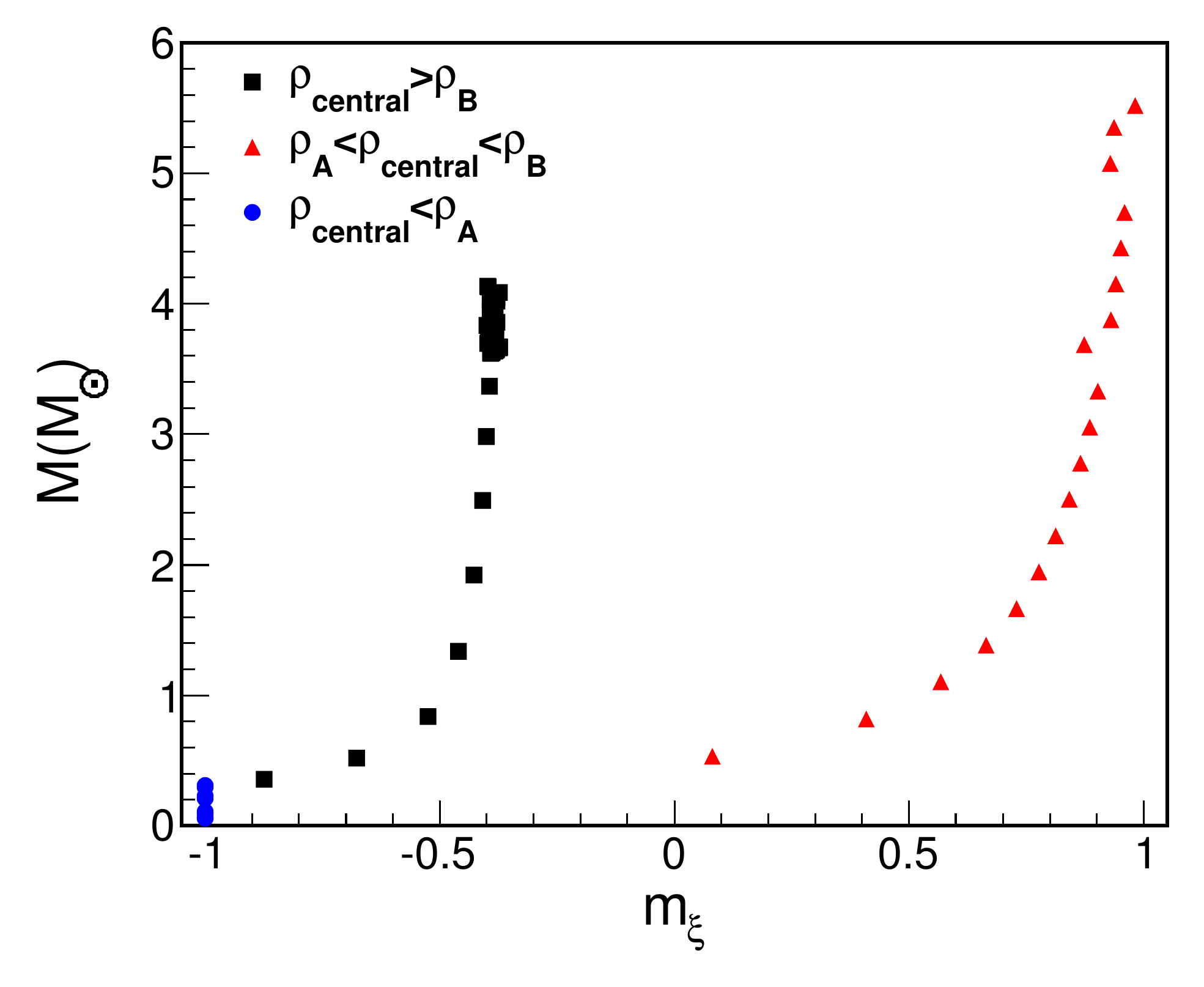}  
\end{tabular}
\caption{(Color online) The total NS mass versus $m_{\xi}$ for different central densities when $K_{sym}=-215$ MeV for CCS$\delta$5.} \label{nfig8}
\end{figure}

For completeness in Fig. \ref{nfig7} we plot the density as function of distance for the same cases discussed in Fig. \ref{nfig6}. Combining these observations it becomes clear that when the stable surface is too small compared to the unstable inner region, then the system might give away some mass to try to reach the conditions of the third panel in the figure. For the unstable cases, the fourth and the fifth panels in Figs. \ref{nfig6} and \ref{nfig7}, the stable region is very small compared to the unstable one and the SN explosions might result.

The question now is if the stable phase A is able to constrain the unstable phase from the center of the star to $R_A$.  We define the (order) parameter $m_{\xi}=\frac{M_{mixedphase}-(M_{phaseA}+M_{phaseB})}{M_{mixedphase}+M_{phaseA}+M_{phaseB}}$, which tells us how much percentage matter is in the unstable phase. In Fig. \ref{nfig8} we plot the total NS mass as function of $m_{\xi}$. The triangle symbols indicate the results obtained with central densities in the unstable region, while the other symbols refer to Fig. \ref{nfig5}, $K_{sym}=-215$ MeV.  From this figure we could classify the NS into two categories according to the $m_\xi$ parameter. In particular if we further assume that stars with $M>2.5M_\odot$ are unstable following the observations (so far), then this will correspond to the conditions $-0.4<m_\xi<-0.35$ and $0.85<m_\xi<1$ roughly. Depending on the speed of expansion in the SN explosion, the matter might have enough time to transform into baryons or more complex nuclei. For very slow explosions the neutrons might decay into protons which then fuse with other neutrons and so on thus forming complex nuclei, similar to the Big Bang nucleosynthesis (BBN). On the other hand if the explosion is too fast, mainly hydrogen will remain. Thus, we could in principle derive the $m_\xi$ parameter from observations using the alternative definition 
$m_\xi'= \frac{M_H-M_{A>H}}{M_H+M_{A>H}}$
where $M_H$ is the mass or density of the ejected hydrogen while $M_{A>H}$ is the mass or density of all the other ions.

\begin{figure} [H]  
\centering
\begin{tabular}{c}
                   \includegraphics[scale=0.5]{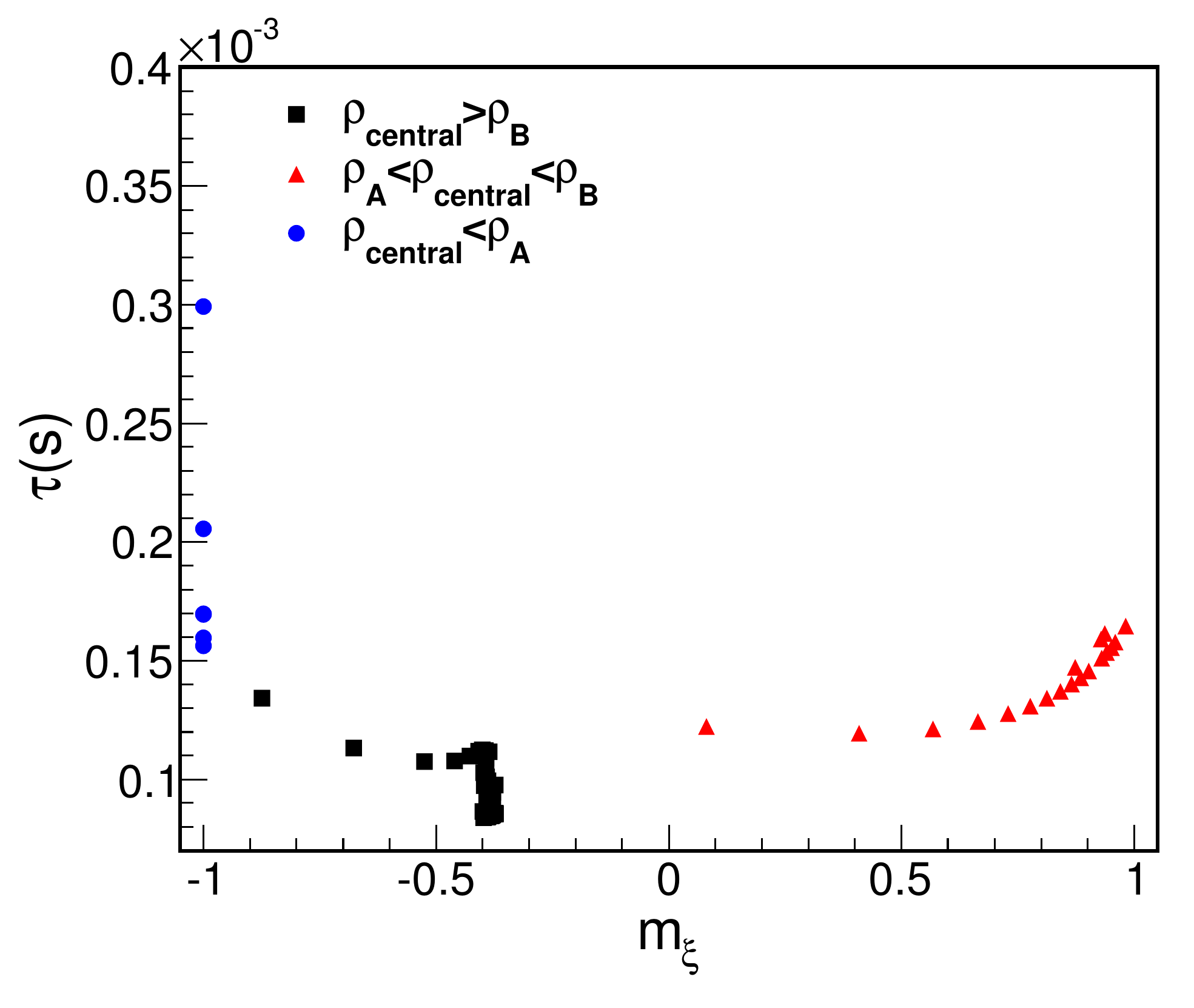}  
\end{tabular}
\caption{The typical time $\tau$ versus $m_\xi$.} \label{nfig9}
\end{figure}

Even though we cannot calculate the conditions when the resulting NS is made mainly of hydrogen or not, we can make some simple estimates. From the radius and the average Fermi velocity (we obtained from the average density), we can define a typical time as: $\tau=R_{NS}/{\bar v_F}$. We plot such a quantity as function of $m_\xi$ in Fig. \ref{nfig9}. Qualitatively we can say that NS having $m_\xi\approx-0.4$ will expand very quickly, thus they might result in SN producing mainly hydrogen. We stress the fact that this will be specially enhanced by other nearby objects which break the spherical symmetry we have assumed. Of course the more massive the nearby object is the faster the explosion might be.

\section{Summary}
In this paper, we have shown that two EoS, i.e. CK225$_1$, CK225 obtained from simple Skyrme interactions and other two, i.e. CCS$\delta$3 and CCS$\delta$5 obtained from Lt, with the same values of $E/A$, $\rho_0$, $K$, $S$, $L$ and $K_{sym}$ at gs density result in different mass-radius relation for neutron stars. We suggest that in order to pin down the EoS we need more constraints at higher densities than the gs one. One constraint might be the critical density (if any) of a first or second order PT. We need to derive the constraints of the EoS from laboratory experiments. We argue that the maximum mass of NS is the results of the competition between the highest power of density in EoS and the PT (if any). 

Fixing $S = 28.5$ MeV, $L=50$ MeV and changing $K_{sym}$ for CCS$\delta$5, we found that it has an instability region and experiences a PT when $K_{sym}= - 215$ MeV. This can be associated to the occurrence of SN phenomena.  For this particular choice of the input parameters, stable NS with very small masses are obtained. These small NS masses are indeed found for all the EoS discussed in this work and others \cite{twomsun, lattimer12, stone2012, huappnp, agrawal2012, lonardoni2014, gandolfi2012, shao2011, logoteta2013, NSmw2, dexheimer2012, svtov1, svtov2, svtov3, svtov4}. These stars are made of neutrons only (no QGP) and their observation might solve the dark matter `puzzle'.  In particular if the PT (either first or second order) is responsible for the SN explosions, then we expect that the resulting average mass distribution (from many SN events) of NS to follow a power mass-law distribution. It is crucial however, to fix to a better precision the values of $S$, $L$ and $K_{sym}$ from experiments in laboratories also using radioactive nuclei.

\section*{Acknowledgments}
J. Sahagun would like to acknowledge the support by DOE, NSF-REU
Program and the support of many people from Texas A\&M Cyclotron Institute. H. Zheng is grateful for the financial support and warm hospitality from INFN-LNS in Italy (where part of the work was performed).

%\section*{References}

\end{document}